# Advantageous punishers in nature

Xinsheng Liu* and Wanlin Guo

*Institute of Nanoscience, Academy of Frontier Science, Nanjing University of Aeronautics and Astronautics, Nanjing 210016, China*

**Abstract.** The evolution and maintenance of cooperation fascinated researchers for several decades. Recently, theoretical models and experimental evidence show that costly punishment may facilitate cooperation in human societies, but may not be used by winners. The puzzle how the costly punishment behaviour evolves can be solved under voluntary participation. Could the punishers emerge if participation is compulsory? Is the punishment inevitably a selfish behaviour or an altruistic behaviour? The motivations behind punishment are still an enigma. Based on public goods interactions, we present a model in which just a certain portion of the public good is divided equally among all members. The other portion is distributed to contributors when paying a second cost. Contributors who are willing to pay a second cost can be costly (and then altruistic) punishers, but they can also flourish or dominate the population, in this case we may call them "advantageous punishers". We argue that most of successful cooperators and punishers in nature are advantageous punishers, and costly punishment mostly happens in humans. This indicates a universal surviving rule: contributing more and gaining more. Our models show theoretically that the original motivation behind punishment is to retrieve deserved payoff from their own contributions, a selfish incentive.

## 1. Introduction

The emergence and abundance of cooperation at various levels in nature poses a challenging and long-standing problem to many disciplines ranging from evolutionary biology to social sciences and economics (1-5). In human societies, costly (and then altruistic) punishment seems an effective mechanism to ensure cooperative behaviour in joint enterprises. People are willing to use altruistic punishment to varying degrees, voluntarily paying a cost to punish noncooperators (6-20). The altruistic punishers can emerge and come to dominate a population of contributors, defectors, and nonparticipants in voluntary public goods games (15-17). If every contributor is willing to pay a cost for punishment, the institutions that impose sanctions on defectors may be formed in the population, and in this case the cooperative behaviour is often enforced more efficiently and the second cost is frequently lower (13). However, in many situations altruistic punishment is costly, the winners don't punish in the sense that costly punishers do not gain the highest total payoff (18, 19). In view of this, whether the costly punishment behaviour emerges if participation is compulsory, whether the punishment behaviour is always costly and altruistic, and what motivates the punishment behaviour are still puzzling. Based on public goods interactions (21-23), representing the prisoner's dilemma of an arbitrary number of players, we present a model in which only a fraction of the public good created from the individual contributions is divided equally among all members. The other part is distributed to contributors when paying a second cost after contribution. The contributors who are willing to pay a second cost are called the persistent cooperators (*PC*), indicating their desire to retrieve the proportion of the payoff derived from their own contributions to the common enterprise with persistent efforts. The persistent cooperators enjoy and engage in the joint enterprise and from it they expect to gain a stable payoff that may not be achieved from one's activities alone. There are many examples that humans share the goods according to their labours. The activities include hunting, warfare, trade, food sharing, and almost all the collective actions that can generate common payoff. Different positions of employment match different levels of salaries, which lays a foundation for modern human societies. People live in highly diverse houses and possess different ranks of cars. In animal societies, for examples, the stronger lions who fight for food share the best meat of buffalo, and in some primates, the dominant animal that has the most powerful fighting ability possesses a lot of priorities (24).

*Email: xsliu@nuaa.edu.cn

The persistent cooperation (*PC*) is a new strategy in our model. The persistent cooperators can behave as costly punishers (*P*) defined by previous models (7, 15-17). But the *PC*s can also fare better within a reasonable range of parameters. We may call the persistent cooperators in this case "advantageous punishers" for comparison, since they behave as punishers (pay another cost to reduce the payoff for defectors through increasing the payoff for themselves), and also they benefit. The inclusion relations for contributors, (pure) cooperators (*C*), persistent cooperators (*PC*), costly punishers and advantageous punishers are shown in Fig. 1. This study focuses on the evolutionary dynamics of a population in which there are three behavioural types: cooperators (*C*), persistent cooperators (*PC*) and defectors (*D*).

## 2. Model and method

Suppose that a large population including cooperators (*C*) and defectors (*D*) has an opportunity to create a public good in the joint enterprise, where cooperators invest into the common good while defectors do not contribute. Cooperators pay an individual cost $c$ to increase the size of the public good by $b$ ($b > c$). The fraction $q$ ($0 < q < 1$) of the public good is divided equally among all members, the other part of it is distributed to contributors when paying a second cost. The second cost is in direct proportion to the fraction of defectors as the punishing cost in previous models (7, 15-17), because the *PC* struggles to retrieve deserved payoff more hardly if there are many defectors in the system. The persistent cooperators (*PC*) are willing to pay the second cost and gain the returning payoff $(1-q)b$, but the cooperators (*C*) who contribute but are not willing to pay a second cost can not obtain the returning payoff.

We first consider a simple situation where all the contributors are willing to pay a second cost. Denote the fraction of the persistent cooperators (*PC*) in the population by $x$, and the second cost by $k(1-x)$, where $1-x$ is the proportion of defectors and $k>0$. Then the expected payoff is $q(bx)-c+(1-q)b-k(1-x)$ $=(b-c)x+[(1-q)b-c-k](1-x)$ for the persistent cooperators, $q(bx)$ for defectors, where $q(bx)$ stands for the expected payoff for every individual derived from the proportion $q$ of the public good and $(1-q)b$ the retrieved portion of payoff for each *PC*. From the two payoff expressions we can see that the game is equivalent to that with pairwise interactions whose payoff matrix is given by

$$\begin{array}{cc} & \begin{array}{cc} PC & D \end{array} \\ \begin{array}{c} PC \\ D \end{array} & \left( \begin{array}{cc} b-c & b(1-q)-c-k \\ qb & 0 \end{array} \right) \end{array} \tag{1}$$

The entries denote the payoff for the row player.

Now we present the typical model in which there are contributors who would not pay a second cost for retrieving the proportion of the payoff. Here there are three behavioral types, cooperators (*C*), persistent cooperators (*PC*) and defectors (*D*). Suppose that the population of size $N$ consists of $n_C$ cooperators, $n_{PC}$ persistent cooperators and $N-n_C-n_{PC}$ defectors. The fractions of three strategies are denoted by $x$, $y$, and $1-x-y$ respectively with $x=n_C/N, y=n_{PC}/N$. The expected payoff for every individual derived from the proportion $q$ of the public good is $qb(x+y)$. The other portion of total common goods is $(1-q)b \cdot (n_C+n_{PC})$, of which the part $(1-q)b \cdot n_{PC}$ derived from the contributions of persistent cooperators is distributed among themselves. Then from this part of payoff each of the persistent cooperators pays a second cost $k(1-x-y)$ for the retrieved payoff $(1-q)b$, but the cooperators do not. If the part $(1-q)b \cdot n_C$ of the public good from the contributions of the cooperators is divided equally among all members, the expected payoff for every individual is $(1-q)bx$, which is given to each of defectors. We consider a more rigorous situation where the persistent cooperators have no intention to obtain the payoff $(1-q)bx$ per individual that should be returned to cooperators (in other situation every persistent cooperator receives an additional share $(1-q)bx$ and they will fare better). Then naturally each of cooperators gets the income $(1-q)bx(n_C+n_{PC})/n_C$, that is, $(1-q)b(x+y)$, from the part $(1-q)b \cdot n_C$ of the total public good derived from the contributions of the cooperators themselves. Thus the expected payoffs for cooperators, the persistent cooperators, and defectors are respectively:

$$P_C = qb(x+y) - c + (1-q)b(x+y) = b(x+y) - c,$$
$$P_{PC} = qb(x+y) - c + (1-q)b - k(1-x-y) = b(x+y) - c + [(1-q)b - k](1-x-y), \quad (2)$$
$$P_D = qb(x+y) + (1-q)bx = b(x+y) - (1-q)by.$$

Thus the equivalent game has the following payoff matrix among C, PC and D

$$\begin{array}{c}  & C & PC & D \\ C & \begin{pmatrix} b-c & b-c & -c \\ PC & b-c & b-c & (1-q)b-k-c \\ D & b & qb & 0 \end{pmatrix} \end{array} \quad (3)$$

The entries denote the payoff for the row player.

## 3. Results

If all the contributors are willing to pay a second cost, then there are only two kinds of players, the persistent cooperators (PC) and defectors (D) in the population. We designate the conditions for the evolution of cooperation in such a system. Let $\alpha = b - c$, $\beta = b(1-q) - c - k$, $\gamma = qb$, and $\delta = 0$ in payoff matrix (1), and then we have the following results (5).

1) PC is an evolutionarily stable strategy (ESS), if $\alpha > \gamma$, that is,
$$\frac{b}{c} > \frac{1}{1-q}.$$

2) PC is risk-dominant (RD), if $\alpha + \beta > \gamma + \delta$. Then we have
$$\frac{b}{c} > \frac{1}{1-q}(1 + \frac{k}{2c}).$$

3) PC is advantageous (AD), indicating that cooperation has a fixation probability greater than the inverse of the population in finite populations, if $\alpha + 2\beta > \gamma + 2\delta$. This induce
$$\frac{b}{c} > \frac{1}{1-q}(1 + \frac{2k}{3c}).$$

4) The persistent cooperators dominate defectors, if $\alpha > \gamma$ and $\beta > \delta$, showing that
$$\frac{b}{c} > \frac{1}{1-q}(1 + \frac{k}{c}).$$

For cooperation to be beneficial, all conditions can be expressed as the benefit-to-cost ratio exceeding a critical value, which is inversely proportional to the parameter $(1-q)$, and proportional to the ratio of the second cost for retrieving the deserved payoff to the contributing cost. Intuitively, if the proportion $(1-q)$ of the payoff that returns to the contributors themselves is large, and the second cost is relatively smaller than the contributing cost, the critical value is small and then the cooperation is easier to prevail.

We next consider the evolutionary dynamics in the typical population in which there are contributors who would not pay a second cost for retrieving the proportion of the payoff. All three behavioral types, cooperators (C), persistent cooperators (PC) and defectors (*D*) are present in the system. From the payoff expressions (2) we can see that if the second cost is high, that is, $k > (1-q)b$, the *PC*s behave as the traditional punishers (*P*) who cooperate and then punish each defector in the population, reducing each defector's payoff by $p/N$ at a cost $k_1/N$ to the punisher, where $p = (1-q)b$ and $k_1 = k - (1-q)b$ (7). In this case the persistent cooperation should be an altruistic act, given that individuals who contribute but do not pay a second cost are better off than the persistent cooperators. As altruistic punishers, the fate of *PC*s has been revealed by previous models. For example, in virtue of models of group selection altruistic punishment is evolutionarily stable when it is common (7, 10). Altruistic punishment can evolve in an individual selection context under the mechanism of voluntary participation, but in a single isolated population it can not emerge (7, 15-17). However, when the second cost is not high, satisfying $k < (1-q)b$, the persistent cooperators fare better than the pure cooperators. In this case the persistent cooperators are advantageous punishers. The magnitude of retrieved payoff $(1-q)b$ comparative to a linear combination of the cooperating cost *c* and the second cost *k* decides the population dynamics. It is reasonable to assume that the cost for cooperating *c* is larger than the cost for retrieved payoff *k* ($c > k$). Then the retrieved payoff $(1-q)b$ is larger than the second cost *k* ($(1-q)b > k$) does not mean it is larger than the cooperating cost *c* ($(1-q)b > c$). But the latter is least condition for cooperation to propagate in a single population without option to abstain from the joint endeavour, just as the elementary condition for the collective activity to be profitable in a model with costly punishment under the mechanism of voluntary participation that the fine of punishment must be larger than the cost of contributing to the public good ($p > c$) (16).

Traditionally, evolutionary game dynamics is described for a single, infinite population. In this case, the standard model of evolutionary selection dynamics is the replicator equations (21, 25, 26). Under replicator dynamics, every strategy that performs better than the population on average increases in abundance $\dot{x}_i = x_i(\pi_i - \bar{\pi})$, where $x_i$ is the fraction of type *i* in the population, $\pi_i$ is the fitness of this type and $\bar{\pi} = \sum_i x_i \pi_i$ is the average payoff in the whole population. Replicator equations describe a deterministic selection process. When the population is finite, the evolutionary game dynamics is often described as an explicit stochastic process which can tackle internal noise arising from the finiteness of the population (27, 28). The stochastic evolutionary dynamics in a finite population is frequently explored under a certain level of selection and mutation (15, 29, 30).

We next specify how strategies diffuse within the population depending on some simple rules that form different ranges of parameters.

(1) If $c < (1-q)b < c + \frac{k}{2}$, then from the payoff matrix (3) we have that both *D* and *PC* are Nash equilibria, but defection is the only one of these three strategies that is an evolutionarily stable strategy (ESS). In this case cooperation is not advantaged.

(2) If $c + \frac{k}{2} < (1-q)b < c + \frac{2k}{3}$, also, both *D* and *PC* are Nash equilibria, and defection is the only strategy that is evolutionarily stable under deterministic selection. However, for stochastic evolutionary dynamics in populations of large but finite size *N*, the *PC* strategy will be favoured in the sense that its abundance (average frequency) exceeds 1/3, the equilibrium abundance of each strategy in the absence of selection, since

$$L_{PC} = \frac{1}{3}\sum_{i=1}^{3}(a_{22} + a_{2i} - a_{i2} - a_{ii}) = \frac{2}{3}\left[(1-q)b - c - \frac{k}{2}\right] > 0.$$

The conclusion is reached for low mutation probability $\mu$ ($\mu \square 1/N$) (29). Interestingly, this range of parameters is the same as that in which persistent cooperators are risk-dominant (RD) in a two-strategy population with only *PC* and *D*.

(3) If $c + \frac{2k}{3} < (1-q)b < c + k$, besides $L_{PC} > 0$, we also have

$$H_{PC} = \frac{1}{3^2}\sum_{i=1}^{3}\sum_{j=1}^{3}(a_{2j} - a_{ij}) = \frac{1}{3}\left[(1-q)b - c - \frac{2k}{3}\right] > 0.$$

Selection favours the *PC* strategy for arbitrary mutation probability (29). Likewise interestingly, in this range the persistent cooperators are advantageous (AD) in a population of only *PC* and *D*. Computer simulations show that the frequency of the persistent cooperators always rises from a smaller value at the beginning, and keeps more than 1/3 most of the time (Fig. 2).

(4) If $(1-q)b > c + k$, then *PC* dominates *D* in a population of only *PC* and *D*. Thus under deterministic selection cooperators may invade an infinitely large population of persistent cooperators in the neutral case, defectors can invade the population of cooperators, and the persistent cooperators can invade the population of defectors. The system may exhibits a tendency to cycle (from cooperation to defection to *PC* and back to cooperation), as a result of a rock-paper-scissors mechanism. Cooperation can be favoured but may not be stabilized (Fig. 3). However, the stochastic evolutionary dynamics in a finite population can show a significant cooperating feature: the persistent cooperators dominate the population almost all the time. As in the models of volunteering and punishment in public goods games, for simplicity we consider the stochastic dynamics under small mutation rates and strong selection (28). This implies that the population is homogeneous most of the time, i.e. in states *PC*, *C* or *D*. An occasional mutant will have taken over the population or disappeared before the next mutation appears (15, 28, 30). Thus the dynamics is determined by a Markov chain based on the transition probabilities between three homogeneous states (the population is composed exclusively of one type of three strategies). In the limit of strong selection, these transition probabilities can be easily derived: (i) in the cooperator state *C*, a single defector is advantageous and takes over the population with probability 1. The persistent cooperators can invade through neutral drift with probability $1/N$; (ii) in the defector state *D*, a mutant cooperator is disadvantageous and disappears. The persistent cooperators are advantageous and take over with probability 1; (iii) in the *PC* state, defectors are disadvantageous. Cooperators obtain the same payoff as persistent cooperators and can take over through neutral drift with probability $1/N$. This yields the transition matrix

$$\begin{array}{c} \\ C \\ D \\ PC \end{array} \begin{array}{ccc} C & D & PC \\ \begin{pmatrix} \frac{1}{2} - \frac{1}{2N} & 0 & \frac{1}{2N} \\ \frac{1}{2} & \frac{1}{2} & 0 \\ \frac{1}{2N} & \frac{1}{2} & 1 - \frac{1}{2N} \end{pmatrix} \end{array}$$

The matrix only depends on the population size *N*. The stationary distribution is

$$P = (P_C, P_D, P_{PC}) = (\frac{1}{N+3}, \frac{1}{N+3}, \frac{N+1}{N+3}).$$

Therefore, for large *N*, the system spends almost all the time in the *PC* state. The reason is that the transition leading away from the *PC* state is neutral and thus very slow compared to all other transitions.

## 4. Discussion and conclusion

Cooperation is an evolutionary puzzle both in human and animal societies. Costly punishment is considered an effective mechanism to ensure cooperation in public goods interactions. This raises a second-order social dilemma because non-punishing cooperators outperform those that do punish (15). Experimental evidence shows that costly punishment does not always increase cooperation, and the winners do not punish. Moreover, if punishers are rare, they suffer tremendous costs from punishing defectors who then undergo little disadvantage.

Under the mechanism of voluntary participation the problem how costly punishment behaviour evolves can be solved (15-17). Besides cooperators, defectors and punishers, these models assume that there exist nonparticipants, the fourth kind of players in the system. The nonparticipants receive a fixed benefit from other activities and play a pivotal role for providing recurrent opportunities for punishers to emerge. However, the costly punishing behaviour can not evolve without any extra mechanism (7, 10, 15-17).

Based on public goods interactions, we present a model in which a certain portion of the public good may return to contributors when paying a second cost. The persistent cooperators who pay a second cost, become the traditional punishers if the second cost is high ($k > (1-q)b$). In this case we mean that the costly punishers emerge in a single population. And the *PC*s are not willing, but compelled to become the costly punishers due to high second cost. In the real world, maybe only human beings can persist in such costly punishment behaviour. Humans have abilities to punish more heavily, and have wisdom to anticipate a stable payoff from the joint enterprise later by persistently imposing sanctions on defectors. Meanwhile the persistent punishing behaviour can benefit other peoples and then be considered an altruistic behaviour (7, 10, 15-17). However, this study has not focused on this situation.

We stress the situation where the second cost is not high ($k < (1-q)b$). In this case the persistent cooperators who are called the advantageous punishers, can flourish or dominate the population, which is just the reason why the contributors are willing to pay a second cost. The possible risk for paying a second cost is that the persistent cooperators may become costly punishers (when the second cost is high in the above mentioned situation). In many practical situations the contributors should pay a second cost to fight against exploitation of defectors and to retrieve deserved payoff from their own contributions, otherwise they can not obtain enough payoffs from cooperation for surviving (see the following example for a typical society of lions and hyenas).

The suggested model in this study has clear biological and social basis. We argue that most of successful cooperators and punishers in the real world are the persistent cooperators and also the advantageous punishers who are not realized before. In this case the social dilemma does not exist, and then cooperation emerge everywhere in nature. In animal societies individuals punish other members that infringe their interests. In most cases, punishers benefit and cooperative behaviours are enforced because victims learn to avoid repeating damaging behaviour (24). These animal punishers are surely advantageous. In a typical society of lions and hyenas, lions pay a first cost to kill buffalos. However, they should pay a second cost to fight hyenas for keeping the meat from robbing (24, 32), and then they are persistent cooperators (*PC*). Usually the second cost is affordable (the number of lions is more than about one-third of the number of hyenas), the lions are advantageous punishers and win. Otherwise they may give up.

In human societies, the persistent cooperators and advantageous punishers are also ubiquitous. For example, most of us work hard to increase social wealth (pay first cost to contribute). Our salaries (the proportion of the payoff derived from our own contributions) are usually the part of the public good generated by ourselves. When we receive the salaries we should simultaneously pay the tax or other fees (the second cost) for public interests or losses. To avoid disadvantage, in many case people usually select to make a warning instead of a fighting, to negotiate rather than to retaliate. We normally settle international affairs and disputes by political or diplomatic means instead of a war. The United States can afford the costly Iraq War, but many of developing countries can not.

To conclude, this model suggests the persistent cooperators (*PC*) who pay a personal cost to increase the size of the public good, and then pay a second cost for the return of partial payoff derived from their own contributions to the common enterprise. The persistent cooperators pursue both the investment and the return. The *PC*s can behave as traditional punishers suggested by previous models, which may account for the origin of human costly punishment behaviour in a population in which participation is compulsory. But they can be favoured within a reasonable range of parameters, in this case they are called advantageous punishers. The occurrence of most punishment behaviours in both animal and human societies is the case. Our model shows theoretically that the original motivation behind punishment behaviour is to retrieve deserved payoff derived from their own contributions, a selfish incentive. The model may also suggest that for most of individuals the major motivation behind punishment and cooperation at various levels is to gain a stable payoff that is maybe not the highest, but seems unattainable without joint efforts, which may account for the nature of life.

**Acknowledgments.** We thank Arne Traulsen for critical comments on the manuscript, and thank W.H. Sandholm and E. Dokumaci for adapting the Dynamo package of Bill Sandholm. This work was supported by National NSF (10971097, 10732040), and the 973 Program (2007CB936204) of China.

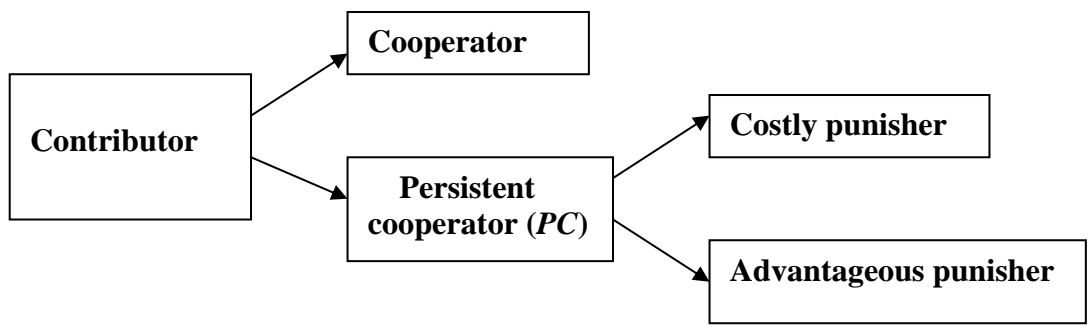

**Figure 1** The inclusion relations for contributors, (pure) cooperators (*C*), persistent cooperators (*PC*), costly punishers and advantageous punishers.

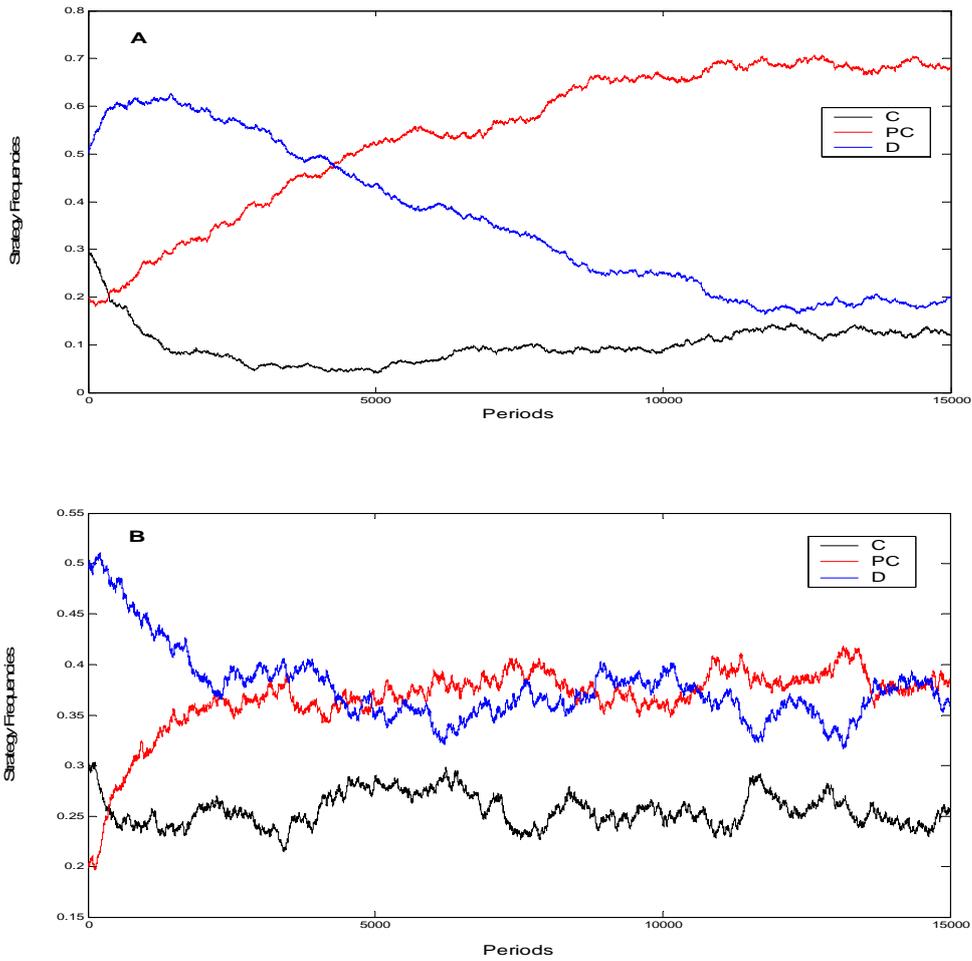

**Figure 2** Simulations of finite populations consisting of three types of players. The selection dynamics are formulated according to the frequency dependent Moran process with population size $N=100$ (Nowak et al. 2004; Traulsen & Hauert 2009). At each time step, an individual is chosen for reproduction proportional to its fitness $f_i$, where $f_i = 1-\omega+\omega P_i$, $P_i$ is the payoff in Equation (2), $\omega$ denotes the intensity of selection (here $\omega = 0.2$ so that all $f_i > 0$), and $i=C$, $PC$, and $D$. One identical offspring is being produced which replaces another randomly chosen individual, but subject to mutation, that is, with probability $\mu > 0$ one of the other two strategies is chosen at random. Initially, the frequency of the persistent cooperators (*PC*) is set to be 0.2. But it quickly rises and keeps more than 1/3 most of the time. The simulation results represent the average frequencies of 50 simulations over 15,000 time periods. Parameter values are $b=3$, $q=1/3$, $c=1.1$ and $k=1$ ($(1-q)b < c+k$). (**A**) The mutation probability is low ($\mu = 0.005 < 0.01 = 1/N$). (**B**) The mutation probability is high ($\mu = 0.05 > 0.01 = 1/N$).

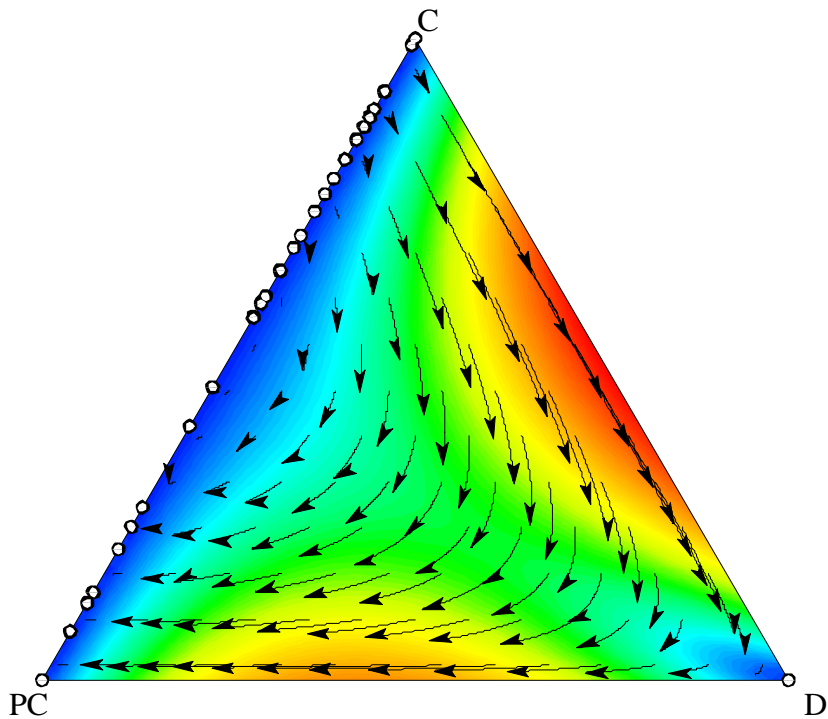

**Figure 3** Replicator dynamics of three strategies depending on the payoff matrix (3) with $b=3$, $q=1/3$, $c=1$ and $k=0.5$ ( $(1-q)b-c-k=0.5>0$ ). Cooperators (*C*) are dominated by defectors (*D*), defectors by the persistent cooperators (*PC*), and cooperators (C) may invade the population of *PC* in the neutral case. There are no stable fixed points in the system. Although the system may exhibits a tendency to cycle, there is a significant region where the population tends toward *PC* (the figure is produced with Bill Sandholm's Dynamo package (Sandholm & Dokumaci 2007)).